\documentclass[12pt]{article}
\usepackage{amsmath,epsfig,graphicx,amssymb}

\newcommand{\beq}{\begin{equation}}
\newcommand{\eeq}{\end{equation}}
\newcommand{\ben}{\begin{eqnarray}}
\newcommand{\een}{\end{eqnarray}}
\newcommand{\bes}{\begin{subequations}}
\newcommand{\ees}{\end{subequations}}
\newcommand{\bFig}{\begin{figure}}
\newcommand{\eFig}{\end{figure}}

\begin{document}

\title{A Simple ``Quantum Interrogation'' Method}
\author{Partha Ghose\footnote{partha.ghose@gmail.com} \\
Centre for Natural Sciences and Philosophy, \\ 1/AF Bidhan Nagar,
Kolkata, 700 064, India\\ and\\Centre for Philosophy and Foundations of Science,\\Darshan Sadan, E-36 Panchshila Park,
 New Delhi 110017, India}

\maketitle
\centerline{\em Date of submission: March 30, 2009; date of acceptance: April 14, 2009}
\begin{abstract}
A simple non-interferometric ``quantum interrogation'' method
is proposed which uses evanescent wave sensing with frustrated total
internal reflection on a surface. The simple method has the advantage over the original interferometric Elitzur-Vaidman method of being able to detect objects that are neither black nor non-diffracting and that are such that they cannot be introduced into an arm of an interferometer for whatever reason (e.g. its size, sensitivity, etc.). The method is intrinsically of high efficiency. 
\end{abstract}

PACS Numbers 03.65 Ta, 42.50 Ct, 42.50 Ex
\section{Introduction}
The basic idea behind ``quantum interrogation'' (or ``interaction-free'' detection as it is often called) is to
detect the presence of an ultra-sensitive object (an object that is
damaged by a single quantum of the probe beam) and also image it without damaging it in most cases. This is of
great significance for imaging photo-sensitive biological systems such as cells and ``delicate'' quantum systems such as Bose-Einstein Condensates, trapped atoms, etc. \cite{white}. 

The best known quantum interrogation method is the Elitzur-Vaidman (EV) method
\cite{elitzur}. In the original version of the method a black object (usually referred to as
the `bomb' which can be triggered by a single incident quantum) is placed in one arm of a balanced Mach-Zehnder interferometer, and the photons arriving at the dark port signal its presence without damaging it. The intrinsic efficiency of the method with 50-50 beam splitters is 33 \% in principle (only $33 \%$
of the bombs can be detected without exploding them). However, the efficiency
can be increased arbitrarily and the restriction to black objects somewhat relaxed by modifying the original method in various ways \cite{kwiat, paul,kwiat2, namekata, gilchrist, mitchison, pavicic}. 

We present here a different quantum interrogation method which does not require an interferometer but uses a combination of the basic ideas of ``evanescent wave sensing'' using ``frustrated total internal reflection'' (FTIR) on a surface. 

\section{Evanescent Wave Sensing with Frustrated Total Internal Reflection}

Consider two adjacent materials with refractive indices $n_i$ and
$n_r$ with $n_i  > n_r$. It is well-known that total internal
reflection (TIR) occurs at the boundary of these materials when
light is incident on the boundary from the material with the higher
refractive index $n_i$ at an angle $\theta_i$ greater than the
critical angle \beq \theta_c = {\rm sin}^{-1}\frac{n_r}{n_i}\eeq The
wave does not, however, vanish in the second medium with refractive
index $n_r$ but is exponentially damped: \beq \psi_{Ev}(x,y) =
\psi(x,0) e^{- y/\xi}\label{pendepth1}\eeq with the `penetration
depth' \beq \xi(n_i, n_r,\lambda) = \frac{\lambda_i}{2\pi \sqrt{n_i^2
{\rm sin}^2 \theta_i - n_r^2}} = \frac{\lambda_i}{2\pi n_i
\sqrt{{\rm sin}^2 \theta_i - {\rm
sin}^2\theta_c}}\label{pendepth2}\,\,,\eeq taking the x-axis along
the boundary and the penetration in the transverse direction y.  The
wave in the second medium with lower refractive index is called the
`evanescent wave'. The electric and magnetic vectors ${\bf E}$ and
${\bf B}$ in an evanescent wave are in time quadrature, and so the
Poynting vector $(c/4\pi ) ({\bf E}\wedge {\bf B})$ vanishes.

If a material with a refractive index $n_t \neq n_r$ comes within
the `penetration depth' $\xi$  of the evanescent wave, it scatters
the wave, i.e. the electric and magnetic vectors are no longer in
time quadrature, a part of the energy leaks (tunnels) out across the
boundary and propagates parallel to the boundary, frustrating total
internal reflection.  Thus, for fixed $\theta_i$ and $n_i$, any
roughness of the surface of the material (variation in y) or
inhomogeneity in its refractive index (variation of $n_r$ along x
due to the presence of the object) will be reflected as intensity
variations in the beam cross-sections.

There are two significant features of evanescent waves in this
context.  Firstly, the component of the momentum perpendicular to
the boundary surface is imaginary, which is why the wave is
exponentially damped and non-radiating.  This implies through
momentum conservation across the boundary that the momentum
components of the evanescent wave parallel to the surface are large.
High momentum components imply small spatial dimensions and high
resolution. Secondly, the expression
(\ref{pendepth2}) for the penetration depth $\xi$  shows that for a
fixed wavelength $\lambda$, the penetration depth increases
indefinitely as $\theta_i$ approaches $\theta_c$. Thus, it is
possible to adjust the penetration depth by varying the angle of
incidence and make it sufficiently large when required. This is
particularly simple when prisms are used as total internal
reflectors.  Optical fibres may be specially modified to take
advantage of this feature.

The above analysis also applies to photons which are quantum
mechanical objects \cite{ghose}. In this case there is tunnelling
across a barrier which is a well-understood quantum
mechanical phenomenon.

\section{The Method}

As shown in the schematic Figure, let a pair of single-photons be emitted from a source $S$ (for example, a pulsed parametric down-conversion source PDC), and let one of the photons be incident on the detector $D_I$ and the other photon be totally internally reflected by the surface $TIR$ into the detector $D_S$. The detection of a photon by $D_I$ can be used to herald the other photon. In the absence of the object and in ideal conditions the two detectors $D_S$ and $D_I$ will count $\bar{n}_0$ photons per second if $\bar{n}_0$ is the flux of signal and idler photons emitted by $S$. However, when an object $O$ with a refractive index different from that of its
surroundings is present on the far side of $TIR$ (as shown in the Figure) but close enough
to its back surface, i.e., within the
`penetration depth' of the evanescent wave which is of the order of
the wavelength of the photons, the evanescent
wave associated with total internal reflection will interact
with the object and a fraction of the photons will tunnel out, partially frustrating total internal reflection. If $\vert \psi\rangle$ is the normalized state of the signal photon before reflection by $TIR$, it splits into an evanescent state and a reflected state after $TIR$ so that the state after $TIR$ can be written as

\begin{figure}[t]\begin{picture}(150,150)(-100,-100) 
\put(1,-20){\line(0,1){90}}
\put(77,-14){\line(0,1){25}}
\put(77,38){\line(0,1){25}}
\put(63,38){\line(1,0){30}}
\put(63,12){\line(1,0){30}}
\put(63,12){\line(0,1){25}}
\put(93,12){\line(0,1){25}}
\put(1,70){\line(1,0){70}}
\put(-15,55){\line(1,1){35}}
\put(1,-20){\line(1,0){70}}
\put(-10,-24){S}
\put(1,-20){\circle*{4}}  
\put(- 40,55){$TIR$} 
\put(72,66){$D_S$}
\put(72,-24){$D_I$}
\put(95,20){$HERALDING\, ARRANGEMENT$}
\put(20,-25){\vector(2,0){10}}
\put(-3,15){\vector(0,2){10}}
\put(25,74){\vector(2,0){10}}
\put(-6,75){\circle*{10}}
\put(-35,82){Object} 
\put(-70,-50){Fig: Quantum Interrogation detection of an object}
\end{picture}\end{figure}

\beq \vert \phi\rangle = 
(a + b) \vert \psi\rangle\label{eq}\eeq with $\vert a \vert^2 + \vert b \vert^2 = 1$ and   
\beq \vert b \vert = {\rm exp} [ - d/\xi(n_i, n_t,\lambda)]\eeq where $d$ is the distance of the object from the back surface of $TIR$ and
$\xi(n_i, n_t,\lambda)$ is given by Eq. (\ref{pendepth2}) above. Thus there is a reduction in amplitude of the state that is totally reflected by $TIR$, and the count rate at $D_S$ is reduced from ${\bar{n}_0}$ to $\vert a \vert^2 {\bar{n}_0}$. Since single photons are sent in one at a time, a photon arriving at $D_S$ could not have `touched' the
object, for then it would have been absorbed by it (assuming the object to be ultra-sensitive or black). Yet the decrease in the counting rate unambiguously indicates the presence of the object. The important point here is that although the photon wave function
splits into an evanescent part and a reflected part at the total
internal reflector (for angles of incidence $\theta_i \geq \theta_c$
), a single photon either interacts with the object and tunnels out,
or it is entirely reflected. These are mutually exclusive
alternatives. This is impossible with multi-photon states of classical
light because with such light reflection and evanescent wave
interaction are concurrent and not mutually exclusive.

Some important differences from the EV method must be pointed out.
In the EV method {\em every} photon in the dark port signals the presence of an ultra-sensitive object. This is not the case here since the detector $D_S$ clicks whether or not an object is present. However, {\em every} photon that is totally internally reflected in the presence of the object and is detected by $D_S$ carries information of its presence through its reduced probability amplitude, resulting in a lower counting rate.

Furthermore, the original EV method works only with a black and non-diffracting object because with a semi-transparent object, for example, the wave function in the arm of the interferometer with the bomb is not fully blocked, and hence interference on the second beam-splitter is not totally destroyed. Consequently, a detection by the dark port cannot obviously be described as completely ``interaction-free''. In the new method even a fully transparent object will allow a fraction of photons to pass through it (so to speak), depending on the gap between the back face of the total internal reflector and the object \cite{mizo}, but these photons leak out and are never detected by $D_S$. Only the photons that are totally internally reflected by $TIR$ are detected and these can be metaphorically said not to ``touch'' the object because if they did, they would have tunnelled out.

However, it is equally important to bear in mind the following similarity between the original EV method and the new method being proposed here. Let us consider the case in which the object to be detected is an ultra-sensitive bomb that is triggered by a single photon. What happens in both cases is that the wave function of the incident photon is split into two parts. Only the method of splitting the wave function is different in the two cases. One part interacts with the wave function of the bomb, and the other part does not. The photons corresponding to the non-interacting part signal the presence of the bomb without exploding it. It is only in this sense that the detection can be said to be ``interaction-free'' in both cases. A better term to use would perhaps be ``damage-free''.

It follows from Eq. (\ref{eq}) that the probability of absorption of a photon by the object is $P_{abs} = \vert b \vert^2$ and the probability of damage-free detection is $P_{dfd} = \vert a \vert^2 \,(a <1)$.
(The limit $\vert b \vert = 0$ and hence $a = 1$ corresponds to the absence of an object within the penetration depth of the evanescent wave, and the method fails.) Hence, for a given object, $P_{dfd}$ can be increased by reducing $P_{abs}$ by placing the object at a suitable distance from the back surface of $TIR$ (the dependence on the distance (Eq. (\ref{pendepth1})) is exponential!). Such flexibility in placing the object makes the method intrinsically of high efficiency, i.e., for any ensemble of identically prepared incident photons, the fraction of photons that interact with the object (triggering the bomb) can be made arbitrarily small compared to the fraction of photons that are totally reflected with a reduced amplitude. 

An advantage of using a single-photon source for damage-free detection (and imaging) over continuous laser light is that the power induced damage is much less in the former case. In a typical cw laser beam the power is of the order of mW to $1$ W and the flux of photons is of the order of $10^{17}$ to $10^{19}$ photons/s whereas single-photon sources have much lower power ($\sim$ pW) and much lower photon fluxes ($10^2$ to $10^6$ photons/s). Although the power in a cw laser can be reduced to levels comparable to a heralded PDC single-photon source by using filters, the statistics is different (Poissonian) from that of single-photon states (sub-Poissonian) \cite{mandel}, the light is still classical \cite{aspect} in character and hence it cannot be used for true damage-free detection. 

\section{Experimental Feasibility}

The feasibility of the method has already been experimentally demonstrated by Mizobuchi and Ohtak\'{e} \cite{mizo} who used a total internal reflector (a prism face) to detect the presence of a second prism within the penetration depth of the incident single photons generated by a Nd:Yag laser, although their results suffered from insufficient statistical precision \cite{unni}. In the present case, a suitable configuration has to be identified to hold the object at controlled distances behind the total internal reflector $TIR$. One way is to follow what Mizobuchi and Ohtak\'{e} did, namely to use Langmuir-Blodgett films. One can also consider using integrated optics methods to fabricate the required $TIR$ as an optical sensor. 

To cut out photons of the original pump laser of frequency $\nu_0$ that are inevitably present as background, suitable interference filters may be used to select the single photons of the right frequency $\nu_1$ to herald the conjugated photons of frequency $\nu_2$ within a coincidence time window of the first detection, together with appropriate logic circuits.

The quantum or {\em shot} noise in a single-photon beam also places certain constraints on the detection of small absorptions by matter over short intervals of time and must be taken care of. Let $\bar{n}_0$ be the mean number of heralded photons received by the detector $D_S$ per second as determined over a sufficiently long time. This is a true population characteristic. Let a sample of $N$ counts be recorded by $D_S$ in $T$ seconds when the object is suspected to be present. By hypothesis the sample mean is then $\mu = \vert a \vert^2 \bar{n}_0$, and assuming the distribution to be sub-Poissonian, the standard deviation $\sigma = \sqrt{\vert a \vert^2 \bar{n}_0\eta}$ where $\eta = 1 - \epsilon \,(\epsilon << 1)$. Then the null hypothesis (namely, that the sample mean $\mu = \bar{n}_0$) can be tested by using the T-test and calculating the statistic  
\ben
t &=& \frac{\mu - \bar{n}_0}{\frac{\sigma}{\sqrt{N}}}\nonumber\\
&=& - \sqrt{\frac{N \bar{n}_0}{\vert a \vert^2 \eta}}\,\vert b \vert^2.\een 
The null hypothesis can be rejected at the $99\%$ confidence level if $\vert t \vert \geq 2.58$ for $N > 30$. If one wishes to restrict the fraction of photons that trigger the bomb to, say $1\%$, then $\vert a \vert^2/\vert b\vert^2 = 10^2$, and 
\beq
N  \geq \frac{6.7243 \times 10^4\,\eta}{\bar{n}_0}.\eeq 
Thus, for example, if $\bar{n}_0 \geq 6.7243 \times 10^2$ photons/s, $N \geq 10^2$ would more than suffice. For small absorptions it would be reasonable to assume that $N \simeq \bar{n}_0 T$, and therefore $T \simeq 0.15$ s would be required to collect a sample size of $100$, and only $1$ bomb would be triggered.

\section{Conclusion}

The method has the advantage over the original EV method of being able to detect objects that are neither black nor non-diffracting and that are such that they cannot be introduced into an arm of an interferometer for whatever reason (e.g. its size, sensitivity, etc.). The object to be detected must, however, be present within the penetration depth of the evanescent wave generated by the sensing surface $TIR$. It is a simple method and is likely to have a wider scope of applications. 

\section{Acknowledgement}
I am grateful to Tathagata Ghose for pointing out to me the appropriateness of the statistical T-test.

\end{document}